\documentclass{ifacconf}

\usepackage{graphicx}      % include this line if your document contains figures
\usepackage{natbib}        % required for bibliography

\usepackage{booktabs}
\usepackage{amsmath}
\usepackage{amsfonts}
\usepackage{algorithm}
\usepackage[noend]{algpseudocode}
\usepackage{algorithmicx}
\usepackage{mathtools}
\usepackage{bm}
\usepackage{tabularx}
\usepackage{color}
\usepackage{soul}
\usepackage{printlen}
\usepackage{pgf}
\usepackage{amssymb}

\newif\ifreviewermode
\reviewermodefalse % uncomment for submission mode

\usepackage[normalem]{ulem}
\newcommand{\remove}[1]{%
  \ifreviewermode
    \textcolor{red}{\sout{#1}}% STRIKETHROUGH
  \else
  \fi
}%
\newcommand{\newtext}[1]{%
  \ifreviewermode
    \textcolor{green}{#1}% GREEN TEXT
  \else
    {#1}%
  \fi
}%

\DeclareMathOperator*{\argmin}{argmin}

\newcommand{\real}{\mathbb{R}}

\algnewcommand{\LineComment}[1]{\State \(\triangleright\) #1}
\algnewcommand\algorithmicswitch{\textbf{switch}}
\algnewcommand\algorithmiccase{\textbf{case}}
\algnewcommand\algorithmicassert{\texttt{assert}}
\algnewcommand\Assert[1]{\State \algorithmicassert(#1)}%

\algdef{SE}[SWITCH]{Switch}{EndSwitch}[1]{\algorithmicswitch\ #1\ \algorithmicdo}{\algorithmicend\ \algorithmicswitch}%
\algdef{SE}[CASE]{Case}{EndCase}[1]{\algorithmiccase\ #1}{\algorithmicend\ \algorithmiccase}%
\algdef{SE}[DOWHILE]{Do}{doWhile}{\algorithmicdo}[1]{\algorithmicwhile\ #1}%
\algtext*{EndSwitch}%
\algtext*{EndCase}%

\begin{document}
\begin{frontmatter}

\title{Integration of a Graph-Based Path Planner and Mixed-Integer MPC for Robot Navigation in Cluttered Environments}

\author[First]{Joshua A. Robbins} 
\author[First]{Stephen J. Harnett} 
\author[First]{Andrew F. Thompson}
\author[First]{Sean Brennan}
\author[First]{Herschel C. Pangborn}

\address[First]{The Pennsylvania State University, 
   University Park, PA 16802 USA (e-mail: jrobbins@psu.edu, sjharnett@psu.edu, thompson@psu.edu, sbrennan@psu.edu, hcpangborn@psu.edu).}

\begin{abstract}                % Abstract of not more than 250 words.
The ability to update a path plan is a required capability for autonomous mobile robots navigating through uncertain environments. This paper proposes a re-planning strategy using a multilayer planning and control framework for cases where the robot's environment is partially known. A medial axis graph-based planner defines a global path plan based on known obstacles, where each edge in the graph corresponds to a unique corridor. A mixed-integer model predictive control (MPC) method detects if a terminal constraint derived from the global plan is infeasible, subject to a non-convex description of the local environment. Infeasibility detection is used to trigger efficient global re-planning via medial axis graph edge deletion. The proposed re-planning strategy is demonstrated experimentally.
\end{abstract}


\end{frontmatter}

\section{Introduction} \label{sec:intro}

A common task for mobile robots is to navigate to a goal position through a cluttered environment while avoiding obstacles. This task is complicated by the possibility that the environment may only be partially known. This paper considers the problem of designing a multilayered planning and control architecture
for robot navigation in partially known, static environments.

At the top level of this architecture, a graph-based planner is used for global navigation through a known map of the environment. Planning for these methods is typically performed prior to routing the path.
Graph search algorithms such as $D^*$~\citep{Stentz1994}, and $D^*$ Lite~\citep{Koenig2005}, are designed to facilitate re-planning as information about the environment is learned. 

One challenge with graph-based planners is that the graph used for planning may be incorrect, especially in uncertain environments. Updating graph connectivity descriptions is, generally, far more computationally expensive than the planning step. To mitigate this challenge, graph-based planners may re-plan by finding alternate routes in the graph without updating the graph.
The $k$-shortest paths algorithm developed by \cite{yen1971finding} is useful for finding a series of shortest paths in a graph. However, these alternates may be highly similar. For example, in a visibility graph formed from an environment of polytopic obstacles, numerous graph edges may traverse the same region of free space. 
\cite{chondrogiannis2020finding} propose a variant of $k$-shortest paths that reduces the similarity of alternate routes by leveraging a path similarity metric.  There is also a variant of the $k$-shortest paths algorithm proposed by \cite{liu2017finding} that works with different similarity metrics. %, as defined by the overlap ratio or the sum of the costs of the shared edges relative to the total length of the path. This is useful in road networks where an edge in the graph typically corresponds to a street, so a diversity of edges corresponds to a diversity of streets.
A blocked corridor can be removed from all of the edges in a graph representation that is built using the medial axis, first described by \cite{blum1967models}, as each corridor has one possible graph edge.  While the medial axis is widely used in path planning~\citep{xu1992motion,masehian2003online,Candeloro2016}, we apply it specifically to the problem of identifying and removing problematic corridors during re-planning.

In many situations, global re-planning can be avoided when navigating through partially known environments by using a multilayer planning architecture wherein a local motion planner is responsible for avoiding unmapped obstacles~\citep{goto1987mobile}. This multilayer approach presents some challenges, however. For instance, sampling-based motion planners may require exhaustive sampling to identify all corridors in an environment~\citep{wang2018learning}, and as such may fail to find a feasible motion plan. Motion planners using convex model predictive control (MPC) formulations can determine if a motion planning problem specification is infeasible with respect to the robot's dynamics and constraints on its motion, but cannot generally consider non-convex obstacle-free spaces. MPC motion planners based on mixed-integer optimization can directly represent a non-convex obstacle-free space at the expense of added computational complexity when compared to convex optimization. See~\cite{ioan2021mixed} for a survey on mixed-integer motion planning.
A limitation of MPC-based motion planners is that they cannot plan outside the local MPC horizon, and the existence of an acceptable motion plan within the horizon cannot be guaranteed \textit{a priori}. In multilayer planning architectures, global re-planning is still necessary to address this limitation.

\subsection{Contributions}

This paper presents a re-planning framework within a multilayered planning and control architecture. The key contributions of this work are: (1) With a mixed-integer motion planner, the objective lower bound is used to determine if a motion planning problem specification is infeasible given a non-convex local map of the environment, (2) an efficient method for updating the global path plan is presented based on edge deletion within a medial axis graph, and (3) the mixed-integer motion planner developed in~\cite{robbins2024mixedintegermpcbasedmotionplanning} is experimentally evaluated in the context of this multilayered control architecture. \remove{This local motion planner was previously evaluated only in simulation and without any coupling to a global path planner.}

\subsection{Outline}
The remainder of this paper is organized as follows. Sec.~\ref{sec:prelims} gives preliminary information including descriptions of mixed-integer MPC and medial axis graph planners, Sec.~\ref{sec:replan} details the re-planning strategy, Sec.~\ref{sec:experiment} presents experimental results, and Sec.~\ref{sec:conclusion} concludes the paper.
\section{Preliminaries} \label{sec:prelims}

\subsection{Notation}
Unless otherwise stated, scalars are denoted by lowercase letters, vectors by boldface lowercase letters, matrices by uppercase letters, and sets by calligraphic letters. Vectors consisting entirely of zeroes and ones are denoted by $\mathbf{0} = \begin{bmatrix} 0 & \cdots & 0 \end{bmatrix}^T$ and $\mathbf{1} = \begin{bmatrix} 1 & \cdots & 1 \end{bmatrix}^T$, respectively. The identity matrix is denoted as $I$. Empty brackets $[\,]$ indicate that a matrix has row or column dimension of zero. Diagonal matrices are denoted as $\text{diag}([\cdot])$. Square brackets following a vector denote an indexing operation such that $\mathbf{x}[i] = x_i$. The Minkowski sum of two sets $\mathcal{A}$ and $\mathcal{B}$ is denoted as $\mathcal{A} \oplus \mathcal{B}$. Expressions using the $\pm$ symbol are expanded using all possible permutations. For instance, $\pm a \pm b \leq c$ expands to the inequalities
\begin{equation}
\begin{matrix}
    a + b \leq c\;,\; -a + b \leq c\;,\; a - b \leq c\;,\; -a - b \leq c\;.\;       
\end{matrix}
\end{equation}

\subsection{Mixed-Integer MPC} \label{sec:MI-MPC-prelim}
Consider the following MPC problem adapted from~\cite{robbins2024mixedintegermpcbasedmotionplanning}:
\begin{subequations} \label{eq:mpc-gen}
\begin{align}
    &\min_{\mathbf{x}, \mathbf{u}} \sum_{k=0}^{N-1} \left[ (\mathbf{x}_k - \mathbf{x}_k^r)^T Q_k (\mathbf{x}_k - \mathbf{x}_k^r) + \mathbf{u}_k^T R_k \mathbf{u}_k \right] \nonumber \\
    & \qquad + (\mathbf{x}_N - \mathbf{x}_N^r)^T Q_N (\mathbf{x}_N - \mathbf{x}_N^r) \;, \label{eq:mpc-gen-cost} \\
    &\text{s.t.} \; \forall k \in \{0, \cdots, N-1 \}\;: \nonumber \\ 
    &\phantom{\text{s.t.}} \; \mathbf{x}_{k+1} = A \mathbf{x}_k + B \mathbf{u}_k\;, \\
    &\phantom{\text{s.t.}} \; \mathbf{y}_{k} = H \mathbf{x}_k,\; \mathbf{y}_{N} = H \mathbf{x}_N \;, \\
    &\phantom{\text{s.t.}} \; \mathbf{x}[0] = \mathbf{x}_0 \;, \mathbf{u}[0] = \mathbf{u}_0 \;, \label{eq:mpc-gen-ic} \\
    &\phantom{\text{s.t.}} \; \mathbf{x}_k \in \mathcal{X},\; \mathbf{x}_N \in \mathcal{X}_N,\;\mathbf{u}_k \in \mathcal{U}\;, \label{eq:mpc-gen-state-input-cons} \\
    &\phantom{\text{s.t.}} \; \mathbf{y}_{k}, \mathbf{y}_{N} \in \mathcal{F} = \bigcup_{i=1}^{n_F} \mathcal{F}_i \subset \real^n \label{eq:mpc-gen-obs-avoid} \;,
\end{align}
\end{subequations}
where $\mathbf{x}_k$ are the system states, $\mathbf{x}^r_k$ are reference states, $\mathbf{u}_k$ are the control inputs, and $\mathbf{y}_k$ are the system outputs. Linear time-invariant (LTI) dynamics are assumed. The sets $\mathcal{X}$, $\mathcal{X}_N$, $\mathcal{U}$, and $\mathcal{F}_i \; \forall i \in \{1, ..., n_F\}$ are convex polytopes. Eq.~\eqref{eq:mpc-gen-obs-avoid} indicates that the constraints on the outputs of the system are in general non-convex. In the context of motion planning, the set $\mathcal{F}$ corresponds to the obstacle-free space. To implement the MPC, \eqref{eq:mpc-gen} is solved over a receding horizon of length $N$, and the optimal state and input trajectories $\{\mathbf{x}_0, ..., \mathbf{x}_N\}$ and $\{\mathbf{u}_0, ..., \mathbf{u}_{N-1}\}$ define a motion plan. 

Eq.~\eqref{eq:mpc-gen} can be formulated as a mixed-integer quadratic program (MIQP) as in~\cite{robbins2024mixedintegermpcbasedmotionplanning}. In that work, $\mathcal{F}$ is represented as a hybrid zonotope~\citep{bird2023hybrid}, though other representations such as  halfspace polytopes and the big-M method are possible~\citep{ioan2021mixed}.

When formulated as an MIQP, \eqref{eq:mpc-gen} can be solved using a branch-and-bound algorithm. Branch-and-bound algorithms can solve mixed-integer convex programs to global optimality by solving a series of convex sub-problems. Key to the operation of these methods is the fact that the optimal objective $j$ of a problem $\pi$ is lower bounded by the optimal objective of any relaxation of that problem. A relaxation $\pi^r \equiv R(\pi)$ is a problem with the same objective function as $\pi$ such that the feasible space of $\pi^r$ is a superset of the feasible space of $\pi$. 

Denote the MIQP representation of \eqref{eq:mpc-gen} as $\pi^{MI}(\mathbf{z})$ where $\mathbf{z}$ are the optimization variables. A general branch-and-bound framework for solving $\pi^{MI}(\mathbf{z})$ is given in Algorithm~\ref{alg:bnb_general}. In this algorithm, $\mathit{CR}(\pi^{MI})$ denotes the convex relaxation of $\pi^{MI}(\mathbf{z})$. A convex relaxation can be constructed by relaxing all integrality constraints in $\pi^{MI}(\mathbf{z})$ to their respective interval hulls, i.e., $z_i \in \{0, 1\} \rightarrow z_i \in [0,1]$. If a problem $\pi^r_i(\mathbf{z})$ is infeasible, then we use the convention that its optimal objective is $j(\pi^r_i) = +\infty$. A branching operation $\texttt{branch}(\pi^r_i)$ creates new sub-problems $\{\pi^r_{j,1}, \pi^r_{j,2} ...\}$ such that $\pi^r_i$ is a relaxation of $\{\pi^r_{j,1}, \pi^r_{j,2} ...\}$. A common method to perform this operation would be to add a constraint that an optimization variable $z_i=0$ for one sub-problem and $z_i=1$ for another. See~\cite{floudas1995nonlinear} for an overview of branch-and-bound methods.

In Algorithm~\ref{alg:bnb_general}, $j_+$ is the objective function upper bound. Similarly, an objective function lower bound can be defined as 
\begin{equation} \label{eq:j-}
j_- =  \min_i j_i^r \;,\; \text{s.t.} \; j_i^r \leq j(\pi^r_i) \;.
\end{equation}
In the branch-and-bound framework, \eqref{eq:j-} can be computed by exploiting the fact that an objective function lower bound is available for each $\pi_i^r \in \bm{\pi}^r$ since a relaxation for each $\pi_i^r \in \bm{\pi}^r$ has been solved. The condition $j_+ - j_- \gg 0$, which checks that the optimization is not converged, is often implemented using a combination of absolute and relative convergence criteria.

\begin{algorithm}[htb]
\textbf{Result}: Optimal solution $\mathbf{z}$ and objective $j$ for the mixed-integer convex program $\pi^{MI}(\mathbf{z})$
\begin{algorithmic}[1]
    \State $(\bm{\pi}^r, \mathbf{z}_+, j_+) \gets (\mathit{CR}(\pi^{MI}), \mathbf{0}, +\infty)$
    \While{$\text{length}(\bm{\pi}^r) > 0$ \textbf{and} $j_+ - j_- \gg 0$}
        \State \textbf{pop} $\pi^r_i$ \textbf{from} $\bm{\pi}^r$ \label{algline:pop-line}
        \State $(\mathbf{z}^r_i, j^r_i) \gets$ \textbf{solve}($\pi^r_i$)
        \If{$j^r_i > j_+$} \textbf{continue}
        \ElsIf{$\mathbf{z}^r_i$ is feasible for $\pi^{MI}(\mathbf{z})$} 
            \If{$j^r_i < j_+$}
                \State $(\mathbf{z}_+, j_+) \gets (\mathbf{z}^r_i, j^r_i)$
                \State \textbf{prune} $\pi^r_i$ s.t. $j^r_i > j_+$ \textbf{from} $\bm{\pi}^r$
            \Else $ $ \textbf{continue}
            \EndIf
        \Else $ $ $\bm{\pi}^r \gets \begin{bmatrix} \bm{\pi}^r & \texttt{branch}(\pi^r_i) \end{bmatrix}$ \label{algline:branch-node}
        \EndIf
    \EndWhile
    \Return $(\mathbf{z}_+, j_+)$
\end{algorithmic}
     \caption{General branch-and-bound framework.}
     \label{alg:bnb_general}
\end{algorithm}

\subsection{Medial Axis Planner}
\label{sec:medial_axis_preliminary}

In an obstacle-filled environment, the medial axis gives the paths with the maximum obstacle clearance. It is defined by the centers of the maximum size disks inscribed in the free space~\citep{choi1999medial} as
\begin{equation}
	\label{eq:medial_axis}
	\mathit{MA}(\Omega) = \{p\in\Omega|B_r(p)\in \mathit{CORE}(\Omega)\} \;,
\end{equation}
where $\Omega$ is the domain of the free space between the obstacles, $B_r(p)$ is the closed disk with radius $r$ centered at point $p$ and $\mathit{CORE}(\Omega)$ is the set of only the maximally inscribed disks.

\subsubsection{Approximation of the Medial Axis}
\label{sec:approximate_medial_axis}

The medial axis can \remove{also }be approximated by Delaunay triangulation of the domain of free space, $\mathit{DT}(\Omega)$~\citep{Dey2004, Delaunay1934}. The circumcenters of triangles who share sides are then connected to find the approximate medial axis. \remove{As an analogy to~\eqref{eq:medial_axis}, }The approximate medial axis, $\mathit{AMA}(\Omega)$, is defined as
\begin{equation}
	\label{eq:approx_medial_axis}
	\mathit{MA}(\Omega) \approx \mathit{AMA}(\Omega) \equiv \{p\in\Omega|Tr(p)\in \mathit{DT}(\Omega)\} \;,
\end{equation}
where $Tr(p)$ is the triangle with circumcenter $p$ and $\mathit{DT}(\Omega)$ is the set of triangles in the Delaunay triangulation of $\Omega$.

\subsubsection{Deriving the Medial Axis Graph}
\label{sec:forming_medial_axis_graph}
Two triangles are neighbors in the triangulation if they share a common side. Triangles with two neighbors are \remove{termed }\textit{2-connected} and triangles with three neighbors are \remove{termed }\textit{3-connected}. \remove{Rather than considering every circumcenter in this triangulation as a node,} For the purpose of creating a searchable graph, only the circumcenters of 3-connected triangles need to be retained as nodes, as these points represent decision points for the planner where the medial axis splits into multiple branches.

Keeping the 3-connected triangles as nodes gives a set of nodes $\mathcal{N}$, defined as
\begin{equation}
    \label{eq:nodes_from_3_conn}
    \mathcal{N} = \{p|Tr(p)\in\mathcal{TR}_3\} \;,
\end{equation}
where $\mathcal{TR}_3$ is the set of 3-connected triangles in $\mathit{DT}(\Omega)$. The series of the circumcenters of the 2-connected triangles between each adjacent pair of 3-connected triangles are noted as the edge between the nodes.
There are two data structures which collectively represent the medial axis graph. 
The first is an adjacency matrix, $A$, defined as
\begin{equation}
    A_{i,j} = \begin{cases}
        &1\quad \text{if node $j$ is connected to node $i$}\;,\\
        &0\quad \text{otherwise}\;.
    \end{cases}
\end{equation}
The second data structure is a mapping indicating the ``triangle chain'' between node $i$ and $j$. This captures the paths between two nodes as an unordered set of all circumcenters sequences that lead from node $i$ to node $j$,
\begin{equation}
    \label{eq:triangle_chain}
    \begin{split}
    \mathit{TC}(n_i,n_j) \mapsto \bigl\{&\{p_i,\dots,p_l,\dots,p_j\},\dots \\
    &\{p_i,\dots, p_k, \dots, p_j\},\dots\bigr\} \;.
    \end{split}
\end{equation}
While the adjacency matrix stores the Boolean indicating that these nodes are connected, the actual medial axis segment representing a curved path between the nodes is stored in triangle chain data structure.

\section{Re-planning Strategy} \label{sec:replan}

\begin{figure*}
    \centering
    \includegraphics[width=0.8\textwidth]{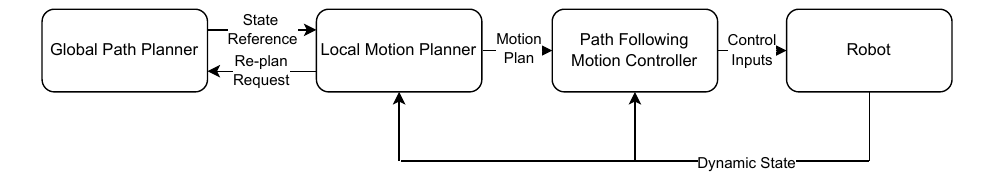}
    \caption{Multilayered planning and control architecture.}
    \label{fig:architecture}
\end{figure*}

In this paper, a medial axis graph-based planner (Sec.~\ref{sec:medial_axis_preliminary}) is used to generate a global path plan for a robot in a cluttered environment. A mixed-integer MPC motion planner (Sect.~\ref{sec:MI-MPC-prelim}) is used to generate local motion plans. \newtext{A diagram of the planning and control architecture is given in Fig.~\ref{fig:architecture}}. To interface the medial axis planner with the MPC, a point on the path plan is used to derive a terminal reference state $\mathbf{x}^r_N$ and a terminal constraint set $\mathcal{X}_N$. Specific implementation details for the robotic system considered in this paper are provided in Sec.~\ref{sec:implementation}.

\subsection{Logic to Trigger Global Re-Planning} \label{sec:replan-mpc}
The mixed-integer MPC motion planner described in Sec.~\ref{sec:MI-MPC-prelim} can detect if a specified terminal constraint $\mathbf{x}_N \in \mathcal{X}_N$ cannot be achieved without violating state, input, or obstacle avoidance constraints subject to the LTI robot dynamics. Branch-and-bound algorithms (i.e., Algorithm~\ref{alg:bnb_general}) can exhaustively determine that a mixed-integer program is infeasible without checking all possible combinations of integer-valued variables. For example, if the convex relaxation $\mathit{CR}(\pi^{MI})$ is infeasible, then $\pi^{MI}$ is determined to be infeasible in only a single mixed-integer iteration. 

In practice, many MPC implementations will use constraint softening~\citep{borrelli2017predictive} to ensure feasibility of the optimization problem. This method uses slack variables to allow for constraint violations, which are penalized severely in the objective function. In the case that softened constraints are used, we define an objective function limit $j_{\text{max}}$ to serve as a proxy for MPC infeasibility. More generally, this limit could be used to determine that the motion plan produced by the MPC is unacceptable. Using branch-and-bound, there is no feasible or acceptable solution to the MPC problem if 
\begin{equation} \label{eq:replan-condition}
j_- > j_{\text{max}} \;,
\end{equation}
where $j_-$ is the objective function lower bound defined in \eqref{eq:j-}. This condition may be met prior to convergence of the branch-and-bound algorithm. In this paper, \eqref{eq:replan-condition} is the condition used to trigger a global re-plan. Once the condition in \eqref{eq:replan-condition} is detected to be true, the branch-and-bound algorithm is terminated.

In a branch-and-bound context, \emph{sharp} mixed-integer representations, i.e., those for which the convex relaxation is the convex hull, are known to result in greater lower bounds $j_-$~\citep{hooker1994logic}. Hybrid zonotopes can be formulated to have sharp relaxations as discussed in \cite{robbins2024mixedintegermpcbasedmotionplanning, glunt2025sharp}. Mixed-integer motion planners using a sharp hybrid zonotope constraint representation can be expected to detect satisfaction of condition~\eqref{eq:replan-condition} in fewer iterations (i.e., fewer sub-problems $\pi^r_i$ solved in Algorithm~\ref{alg:bnb_general}) than motion planners using representations that do not have this property, such as those based on the big-M method~\citep{hooker1994logic}. This paper uses the hybrid zonotope-based MPC motion planning problem formulation described in~\cite{robbins2024mixedintegermpcbasedmotionplanning}.

\subsection{Global Re-Planning Algorithm}
When condition~\eqref{eq:replan-condition} is met, edge deletion is used to remove the corridor that cannot be traversed from the medial axis graph.
From the vehicle's current position, $q_{\text{veh}}$, the edge the vehicle is currently routing can be identified. 
The circumcenter nearest the vehicle's current position is
\begin{equation}
    \label{eq:p_near}
    p_{\text{near}} = \underset{p \in \mathit{AMA}(\Omega)}{\argmin} ||p - q_{\text{veh}} || \;.
\end{equation}
As the medial axis defines the free-space positions farthest from obstacles, the nearest circumcenter to the vehicle must be in the same corridor as the vehicle.
\remove{i.e., the nearest circumcenter to the vehicle will never be on the opposite side of an obstacle from the vehicle's position.} Algorithm~\ref{alg:replan} gives the procedure for removing this corridor.

\begin{algorithm}[ht]
	\caption{\strut Remove the currently occupied corridor from the medial axis graph.
    \remove{Algorithm for removing the currently occupied corridor from a medial axis graph for generating a new global path plan without reusing this corridor.}\\ \textbf{Input:} $A$, adjacency matrix; $TC$, function mapping triangle chains to nodes and its range, $\mathcal{R}(TC)$; $p_{\text{near}}$, nearest triangle circumcenter to the vehicle; $R$, current global route as an ordered series of triangle chains.\\ \textbf{Output:} $A^{new}$, adjacency matrix with current corridor removed; $TC^{new}$, function mapping triangle chains to nodes with current corridor removed.}\label{alg:replan}
	\begin{algorithmic}[1]
		\State $A^{new} \gets A$ %\Comment{Copy the medial axis graph}
		\State $TC^{new} \gets TC$
        \State $tc_{near} \gets R \cap \{tc \in \mathcal{R}(TC) | p_{\text{near}} \in tc\}$ %\Comment{Identify the current graph edge (i.e., corridor) as the corridor in the route that contains the nearest circumcenter.}
        \State $\mathcal{R}(TC^{new}) \gets \mathcal{R}(TC) \setminus tc_{near}$ %\Comment{Remove this corridor from the triangle chain mapping.}
        \For{$\{\forall (n_i,n_j) \in \mathcal{N} | TC(n_i,n_j) \mapsto \emptyset\}$}
            \State $A^{new}_{n_i,n_j} \gets 0$ %\Comment{If removing that edge results in two nodes no longer having an edge between them, indicate this in the adjacency matrix.}
        \EndFor
	\end{algorithmic}
\end{algorithm}	

A new edge is added along the medial axis, from $q_{\text{veh}}$ to the last node successfully routed through. This allows backtracking out of the current corridor when rerouting. From this state of the graph, the search algorithm can be called again to return the best path that excludes the current corridor. Because the medial axis is never fully recalculated and no triangulation of free space is required, this procedure scales in linear time; the only step\remove{ in this algorithm} where graph size influences \remove{the }run time is \remove{when }identifying the edge to be removed based on vehicle position.
\section{Experimental Evaluation} \label{sec:experiment}
This section describes an experimental evaluation of the planning and control strategies described in this paper. Algorithms were implemented using a combination of C++, Python, and MATLAB within a ROS2 framework on an Ubuntu 22.04 desktop with an Intel Core i7-14700 processor and 32~GB of RAM. The test platform was a Husarion ROSbot 2R, which is a lab-scale differential drive robot. Optitrack Prime$^\text{X}$22 motion capture cameras were used to provide state feedback.

\subsection{Implementation} \label{sec:implementation}
\begin{figure*}
    \centering
    \input{figs/replan_example/experiment_visualization.pgf}
    \caption{Experimental demonstration of global re-planning triggered by condition~\eqref{eq:replan-condition} being satisfied during solution of the MPC mixed-integer optimization problem. \newtext{Video: https://www.youtube.com/watch?v=sZKpWM2NG9I.}
    }
    \label{fig:replan-experiment}
\end{figure*}

\subsubsection{Robot Dynamics Model}
A unicycle model with first-order speed and turn rate dynamics is used to model the motion dynamics of the ROSbot 2R, i.e., 
\begin{subequations} \label{eq:robot-dyn}
\begin{align} 
    &\dot{x} = v \cos{(\theta)} \;,\; \dot{y} = v \sin{(\theta)} \;,\; \dot{\theta} = \omega \;, \label{eq:robot-dyn-unicycle} \\
    &\dot{v} = \frac{1}{\tau_v} (v_r-v) \;,\; \dot{\omega} = \frac{1}{\tau_\omega} (\omega_r-\omega) \;,
\end{align}
\end{subequations}
where $x$ and $y$ are position coordinates, $\theta$ is the heading angle, $v$ is the linear speed, and $\omega$ is the turn rate. The system time constants $\tau_v$ and $\tau_{\omega}$ were estimated to be $0.2$~s and $0.3$~s, respectively, using system identification techniques.
The speed and turn rate set-points are $v_r$ and $\omega_r$. Unicycle models accurately model differential drive robots~\citep{becker2014controlling}.

\subsubsection{Path-Following Motion Controller} \label{sec:path-following-controller}
A path-following motion controller is used to track the motion plan generated by the mixed-integer MPC (Sec.~\ref{sec:MI-MPC-prelim}) subject to \eqref{eq:robot-dyn}. This controller consists of the control laws
\begin{subequations} \label{eq:path-following-control-laws}
\begin{align}
    &v_{\text{cmd}} = k_t e_t \cos{(\theta_r-\theta)} + v_r \;, \label{eq:path-following-tangential} \\
    &\theta_{\text{cmd}} = \arctan{\left(\frac{k_n e_n}{v}\right)} + \theta_r \;, \label{eq:path-following-normal} \\
    &\omega_{\text{cmd}} = k_{\theta} (\theta_{\text{cmd}} - \theta) + \omega_r \;,
\end{align}
\end{subequations}
where $e_t$ is the position error component tangent to the specified path, and $e_n$ is the component normal to the path. The reference velocity $v_r$, heading $\theta_r$, and turn rate $\omega_r$ come from the MPC motion planner and are used for feedforward control. The $\cos{(e_{\theta})}$ factor in~\eqref{eq:path-following-tangential} and the $\arctan{(\cdot)}$ operation in~\eqref{eq:path-following-normal} are geometric corrections. The tangential position gain is $k_t=1.5$~1/s, the normal position gain is $k_n=0.6$~rad$\cdot$s, and the heading gain is $k_{\theta}=1.9$~1/s.

\subsubsection{MPC Motion Planner}
The MPC motion planner is based on the unicycle dynamics model~\eqref{eq:robot-dyn-unicycle}. The unicycle model is differentially flat in terms of the position states $x$ and $y$ as described in~\cite{diffflat}, and permits motion planning using the discrete time LTI double integrator model
\begin{equation} \label{eq:dbl-int}
    \begin{bmatrix}
        x_{k+1} \\ y_{k+1} \\ \dot{x}_{k+1} \\ \dot{y}_{k+1}
    \end{bmatrix} = \begin{bmatrix}
    1 & 0 & \Delta t & 0 \\
    0 & 1 & 0 & \Delta t \\
    0 & 0 & 1 & 0 \\
    0 & 0 & 0 & 1
\end{bmatrix} \begin{bmatrix}
        x_{k} \\ y_{k} \\ \dot{x}_k \\ \dot{y}_{k}
\end{bmatrix} + 
\begin{bmatrix} \frac{1}{2} \Delta t^2 & 0 \\ 0 & \frac{1}{2} \Delta t^2 \\ \Delta t & 0 \\ 0 & \Delta t\end{bmatrix} \begin{bmatrix}
    \ddot{x}_k \\ \ddot{y}_k
\end{bmatrix} \;,
\end{equation}
where $k$ is the discrete time step. Optimal trajectories of the double integrator system \eqref{eq:dbl-int} are transformed into unicycle model states as
\begin{equation}
    v_r = \sqrt{\dot{x}^2 + \dot{y}^2} \;,\; \theta_r = \text{atan2}(\dot{y},\; \dot{x}) \;,\; \omega_r = \frac{\dot{x} \ddot{y} - \dot{y} \ddot{x}}{\dot{x}^2 + \dot{y}^2} \;.
\end{equation}

As described in~\cite{whitaker2021optimal}, polytopic approximations of velocity and turn rate constraints are
\begin{equation} \label{eq:dbl-int-cons}
    \pm \dot{x}_k \pm \dot{y}_k \leq v_{\text{max}} \;,\; \pm \ddot{x}_k \pm \ddot{y}_k \leq v_{\text{min}} \omega_{\text{max}} \;,
\end{equation}
where $v_{\text{max}}$ and $v_{\text{min}}$ are maximum and minimum velocities, respectively, and $\omega_{\text{max}}$ is the maximum turn rate. In this MPC implementation, $v_{\text{max}}=0.5$~m/s, $v_{\text{min}}=0.1$~m/s, and $\omega_{\text{max}}=\pi$~rad/s. Eq.~\eqref{eq:dbl-int-cons} is used to define $\mathcal{X}$ and $\mathcal{U}$ in~\eqref{eq:mpc-gen}. 

The terminal constraint set $\mathcal{X}_N$ is given as
\begin{equation} \label{eq:term-cst-set}
    \mathcal{X}_N = \left\{ \begin{bmatrix}
       x & y & \dot{x} & \dot{y}
    \end{bmatrix}^T \middle| \begin{bmatrix}
        x \\ y
    \end{bmatrix} \in  \begin{bmatrix}
        x^r_N \\ y^r_N
    \end{bmatrix} \oplus \mathcal{P}_N \;,\; \begin{bmatrix}
        \dot{x} \\ \dot{y}
    \end{bmatrix} = \mathbf{0} \right\} \;,
\end{equation}
where $x^r_N$ and $y^r_N$ are reference positions along the global path plan. These are computed using a fixed lookahead distance as in line-of-sight guidance~\citep{fossen2021handbook}, which is set to 2~m in this case. The set $\mathcal{P}_N$ defines how much the terminal position may deviate from the reference position. In this paper, $\mathcal{P}_N$ is a regular hexagon represented as a zonotope. The velocity is required to be zero at the end of the MPC horizon $N$ to ensure persistent feasibility.

A local map of the environment is used to generate the obstacle-free space set $\mathcal{F}$, which defines the obstacle avoidance constraints in~\eqref{eq:mpc-gen}. Obstacles in the local map are bloated to account for inter-sample constraint violations, and a convex partition is constructed using the Hertel and Mehlhorn algorithm~\citep{o1998computational}. This partition is transformed into a hybrid zonotope using Thm.~5 of~\cite{siefert2025reachability}. In this paper, the local map boundary is an axis-aligned box with 2.1~m width and height.

All constraints except for those on the control inputs $\ddot{x}_k$ and $\ddot{y}_k$ are subject to constraint softening with a quadratic cost of $1e6 \cdot s_i^2$ for each slack variable $s_i$. The maximum cost for triggering global re-planning is set to $j_{\text{max}}=1000$. \newtext{This parameter was selected as a proxy for infeasibility given the cost function, MPC horizon, and lookahead distance. In other words, $j_-$ should never exceed $j_{\text{max}}$ if the motion planning problem is feasible.}

Referencing \eqref{eq:mpc-gen}, the MPC cost function is defined by the matrices
\begin{subequations}
\begin{align}
    &Q_k = \text{diag}(\begin{bmatrix}
        0.1 & 0.1 & 0 & 0
    \end{bmatrix}) \;, \\
    &R_k = \text{diag}(\begin{bmatrix}
        10.0 & 10.0
    \end{bmatrix}) \;, \\
    &Q_N = \text{diag}(\begin{bmatrix}
        10.0 & 10.0 & 0 & 0
    \end{bmatrix}) \;.
\end{align}
\end{subequations}
The MPC horizon is $N=15$ and the discrete time step is $\Delta t=0.5$~s. The reference state $\mathbf{x}^r_k \; \forall k \in \{0, ..., N-1\}$ is set equal to $\mathbf{x}^r_N$.

\subsubsection{Global Map Generation}
For this work, a random, virtual obstacle map is generated. This map consists of polytopic obstacles where the larger obstacles are assumed to be mapped and therefore known to the global path planner, while smaller obstacles are unmapped. The local map used by the motion planner accounts for both small and large obstacles and would be generated using data from onboard sensors in practice.

\subsection{Experimental Results} \label{sec:stack-simulation}
This section presents experimental results for the planning and control algorithms described in this paper. Fig.~\ref{fig:replan-experiment} depicts several key snapshots of a representative scenario. The laboratory and robot are captured with an overhead \remove{fisheye }camera, and virtual obstacles and planning information are overlaid on top of this image.\remove{ Note that due to imperfect correction of the fisheye effect, the robot position in the image occasionally deviates slightly from the overlaid trajectory.} 

In Fig.~\ref{fig:replan-experiment}(a), there is no dynamically feasible trajectory that avoids obstacles and satisfies the terminal constraint. The mixed-integer MPC motion planner detects satisfaction of the re-planning condition \eqref{eq:replan-condition} after 474 mixed-integer iterations with $j_- = 1118.6$, and an update to the global path plan is requested. The path planner takes 28~ms to re-plan, and the reference state $\mathbf{x}^r_N$ and terminal constraint $\mathcal{X}_N$ are updated. Fig.~\ref{fig:replan-experiment}(b) shows that the motion planning problem specification is still infeasible, so the re-plan condition is again triggered after 273 iterations with $j_- = 1120.9$. The path planner took 5~ms to re-plan in this case. In both Figs.~\ref{fig:replan-experiment}(a) and \ref{fig:replan-experiment}(b), the depicted motion plan is the last motion plan prior to \eqref{eq:replan-condition} becoming true. In Fig.~\ref{fig:replan-experiment}(c), a second updated path has been received, and the motion planner finds a dynamically feasible trajectory that satisfies all constraints. 
Fig.~\ref{fig:replan-experiment}(d) shows the final result of the experiment: The robot reaches its destination while avoiding mapped and unmapped obstacles.

As discussed in Sec.~\ref{sec:intro}, multilayer planning architectures that rely on MPC to avoid unmapped obstacles may fail when no acceptable motion plan exists within the MPC horizon. The experimental results demonstrate that this challenge can be mitigated by using condition~\eqref{eq:replan-condition} to efficiently remove corridors in a global, graph-based planner.

\section{Conclusion} \label{sec:conclusion}
A multilayered planning and control architecture was developed to navigate a robot through a cluttered, partially known environment. A medial axis global planner is responsible for finding a path that avoids mapped obstacles, while a mixed-integer motion planner is responsible for generating dynamically feasible trajectories that avoid both mapped and unmapped obstacles. By leveraging information computed as part of an optimization routine within the motion planner, infeasible motion planning problem specifications--or those for which no acceptable solution exists--can be detected. This information is used for efficient global re-planning via edge deletion. Experimental results demonstrate the efficacy of the proposed approach.

\bibliography{bibitems}             % bib file to produce the bibliography
                                                   
\end{document}